\documentclass[useAMS,usenatbib]{mn2e}

\usepackage{times}
\usepackage{graphicx}

\def\dis{\displaystyle}
\def\note #1]{\newline{\bf $\rightarrow$ #1]}\newline}

\def\eg{{e.g.}}
\def\etal{{et~al.}}
\def\HeII{He{\sc ii} }

\def\Ad{A_{\rm d}}

\def\btaud{\bar\tau_{\rm d}}
\def\del{\triangle}
\def\delc{\triangle_{\rm c}}
\def\deld{\triangle_{\rm d}}
\def\delg{\triangle_g}
\def\delr{\triangle_\rho}
\def\deltad{\delta_{\rm d}}
\def\devdf#1#2{{\diff #1 \over \diff #2}}
\def\devds#1#2{{\diff^2 #1 \over \diff #2^2}}
\def\dgamma{{\delta\Gamma_1 \over \Gamma_1}}
\def\dgammap{{\delta(\Gamma_1\! P) \over (\Gamma_1\! P)}}
\def\diff{{\rm d}}

\def\drho{{\delta\rho \over \rho}}
\def\Lambdad{\Lambda_{\rm d}}
\def\ltil{\tilde{l}}
\def\muHz{\, \mu{\rm Hz}}
\def\omegat{\tilde\omega}
\def\taud{\tau_{\rm d}}
\def\taut{\tau_{\rm t}}

\def\infig#1{
   \centerline{\includegraphics[width=\hsize]{\figdir/#1} } }
\def\figdir{./}

\begin{document} 
 
\title[Second ionization region of helium in the Sun.
       I. Sensitivity study and methodology]
      {Seismic analysis of the second ionization region of 
       helium in the Sun: 
       I. Sensitivity study and methodology}

\author[M\'ario~J.~P~.F.~G.~Monteiro \& Michael~J.~Thompson]
       {M\'ario~J.~P.~F.~G.~Monteiro$^{1}$\thanks{E-mail:
           mjm@astro.up.pt; Michael.Thompson@sheffield.ac.uk}$^{{,}2}$
        and
        Michael~J.~Thompson$^{3}$\footnotemark[1]\\
   $^1$ Centro de Astrof\'{\i}sica da Universidade do Porto,
        Rua das Estrelas, 4150-762 Porto, Portugal\\
   $^2$ Departamento de Matem\'atica Aplicada,
        Faculdade de Ci\^encias da Universidade do Porto,
        Rua do Campo Alegre 687, 4169-007 Porto, Portugal\\
   $^3$ Department of Applied Mathematics, University of Sheffield, 
        Sheffield S3 7RH, UK}

\date{Accepted 2005;  Received 2005; in original form 1996}
\pagerange{\pageref{firstpage} -- \pageref{lastpage}}
\pubyear{2005}

\maketitle
\label{firstpage}
\begin{abstract}
  The region of the second ionization of helium in the Sun is a narrow
layer near the surface.
  Ionization induces a local change of the adiabatic exponent $\Gamma_1$,
which produces a characteristic signature in the frequencies of p-modes.
  By adapting the method developed by \citet*{monteiro94},
we propose a methodology for determining the properties of
this region by studying such a signature in the frequencies of oscillation.

  Using solar data we illustrate how the signal from the helium
ionization zone can be isolated.
  Using solar models with different physics -- theory of convection,
equation of state and low temperature opacities --
we establish how the characteristics of
the signal depend on the different aspects contributing to the structure
in the ionization layer.
  We further discuss how the method can be used to measure the
solar helium abundance in the envelope and to constrain the physics
affecting this region of the Sun.

The potential usefulness of the method we propose is shown. It may
complement other inversion methods developed to study the solar structure and
to determine the envelope helium abundance.
 \end{abstract}


\section{Introduction}\label{sec:int}

  The direct determination of the helium abundance in the solar near-surface
layers is difficult and uncertain, 
although it is very important to
the modelling of the internal structure and evolution of the Sun
(see \citealt{kosovichev92} for a comprehensive
discussion of the sources of uncertainties).
  It is usually taken as a fitting parameter of an evolutionary sequence
that provides the correct luminosity for the Sun at the present age.
  The possibility of constraining this parameter to have the observed
value for the Sun is important to improve the mass loss estimates and early
evolution of the Sun, as well as to test the effects of mixing and
settling on stellar evolution.
 
  Several attempts have been made to use solar seismic data to calculate
the abundance of helium ($Y$) in the solar envelope
\citep*{dziembowski91,vorontsov91,vorontsov92,cd92b,hernandez94,antia94,basu95,gough95,richard98}.
  However the dependence of the determination on other aspects, in
particular the equation of state, poses serious difficulties to an
accurate direct seismic measurement of the envelope abundance of helium
\citep{kosovichev92,hernandez94,basu97}.
  The sensitivity of the modes to the helium abundance is primarily provided
by the change of the local adiabatic sound speed $c$ due to ionization.
  Such sensitivity is given by the behaviour of the first adiabatic exponent,
$\Gamma_1$, since $c^2 {\equiv} \Gamma_1 p/\rho$ where $p$ and $\rho$ are the 
pressure and density respectively,
and consequently it is strongly dependent on the assumed equation of state
and other physics relevant for the region where the ionization takes
place.
  This is the main reason why the seismic determination of the envelope
abundance of helium is highly complex.

  Here we propose a method complementary to those used previously, 
by adapting the procedure
developed by \citet[ in the following MCDT]{monteiro94}
 and \citet*[ in the following CDMT]{cd95}.
  In using the solar frequencies in a different way, which provides a
direct probe to the region of ionization, we aim at providing a method
where the different effects at play in the ionization zone can be
isolated, constructing a procedure to access the chemical abundance.
  Localized variations in the structure of the Sun, such as occur
at the base of the convective envelope (see MCDT and \citealt*{monteiro96t})
and in the region of the second ionization of helium \citep{monteiro96t},
create a characteristic signal in the frequencies of oscillation.
   The properties of such a signal, as measured from the observed
frequencies, are related to the location and thermodynamic properties of
the Sun at the layer where the sharp or localized variation occurs.
   The main advantage we see in this method is the possibility to 
utilise different characteristics of the signal to distinguish different 
aspects of the
physics of the plasma at the region where the signal is generated.
   In particular we may be able to separate the effects due to convection,
the low-temperature opacities and the equation of state from the 
quantification of the helium abundance that we seek to achieve.
  Here we mainly concentrate on separating these different contributions
in order to establish the dependence of the parameters of the signal in
the frequencies on the different aspects of the structure at the
ionization region.
   Using a variational principle we determined how the zone of the
second ionization of helium can indeed be considered as a localized
perturbation to an otherwise `smooth' structure, generating a
characteristic signal in the frequencies of the modes.

   We note that simplified versions of the expression for the signal
discussed here have been applied successfully to cases where there
are only very low degree frequencies.
The signal has been fitted either to the frequencies of low degree modes
\citep*{monteiro98,verner04}, or to frequency differences
\citep*{miglio03,basu04,vauclair04,bazot04,piau05}.
   Here we obtain the expression for the general case of having also 
modes of higher
degree, of which the low degree applications are a particular case.
   We also demonstrate the method for making use of the information
in moderate-degree data available only for the Sun.
   When using modes with degree above 4 or 5 we can avoid using frequencies
affected by the base of the convection zone and may hope to achieve a much
higher precision in the results as many more frequencies with lower
uncertainties can be used.

   In this work we present the analysis of the characteristics of the
signal under different conditions.
   Several models with different physics and envelope helium abundances
are used to test the method in order to prepare the application to the
observed solar data.

\section{The region of the second ionization of
         helium}\label{sec:hel_zone}

  In order to model the sensitivity of the modes to this region we must
try first to understand how ionization changes the structure.
   Secondly, we need to estimate how the modes are affected by such a
region.
  The details of the derivations are discussed in the Appendix, but the
assumptions and the main expressions are reviewed and analysed here.

\subsection{Properties of the ionization region}

  Because the helium second ionization zone ({\HeII} ionization zone)
is sufficiently deep (well within the
oscillatory region of most of the modes) we propose to adapt the method
discussed in MCDT to the study of this layer.
   The contribution from a sharp variation in the structure of the Sun to
the frequencies can be estimated by calculating from a variational
principle for the modes the effect of a localized feature.
   In the work by MCDT the feature was the base of the convection zone and
the sharp transition was represented by discontinuities in the derivatives
of the sound speed.
   Because of the size of the ionization region when compared with the
local wavelength of the modes, that representation is not adequate to
reproduce the effect on the frequencies for the ionization region.

\begin{figure}
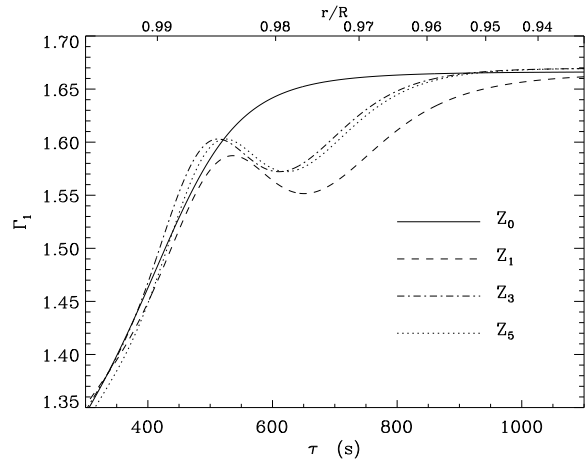

\centering \infig{fig01}
 \caption{Plot of the adiabatic exponent $\Gamma_1$ for several solar
models.
   As a reference we have calculated a model ($Z_0$) where the second
ionization of helium has been suppressed.
   The other two models are calculated using different equations of state
(see Table~1 for further details on the models).
   The second ionization of helium takes place around an acoustic depth of
600~s, corresponding to the depression on the value of $\Gamma_1$. }
 \label{fig:gamma1}
\end{figure}

   Here we must, instead, consider how the ionization changes the
first adiabatic exponent $\Gamma_1 {\equiv} (\partial\ln p/\partial\ln\rho)_s$ 
(the derivative at constant specific entropy $s$) locally, 
generating what can be described
as a `bump' over a region of acoustic thickness of about 300~s (see
Fig.~\ref{fig:gamma1}).
   This allows us to estimate how the frequencies of oscillation are
`changed' due to the presence of this feature in the structure of the Sun.
   The effect will be mainly taken into account through the changes induced
in the adiabatic gradient $\Gamma_1$ by the ionization.
   Other thermodynamic quantities are also affected, but the changes on
the local sound speed is mainly determined by changes in $\Gamma_1$.
   Therefore, we will calculate the dominant contribution to the changes
in the frequencies by establishing what is the effect on the modes due to
changes of the adiabatic exponent.

  \cite{dappen86} and \cite*{dappen88}
have proposed a method based on the same principle,
by using the sensitivity of the sound speed to changes on the adiabatic
exponent.
   Using this sensitivity they calibrate a quantity that is associated
with ionization in order to try to measure the helium abundance in the
solar envelope from seismic data.
   But most methods have difficulties in removing the dependence of the
calibration on the physics of the reference models, making it difficult to
obtain a value for the abundance.
   This is the problem we try to address in this contribution by
developing a method able to measure in the frequencies the effect of the
ionization and its dependence on the abundance, isolated as much as
possible from the other uncertainties.

\subsection{A variational principle for the effect on the frequencies}

  A variational principle for nonradial adiabatic oscillations, assuming
zero pressure at the surface located at radius $R$ as a boundary
condition, can be formulated.
   It is possible to further consider only higher-order acoustic modes, for
which we may neglect the perturbation in the gravitational potential.
   The outcome of such a variational principle is an estimate of how the
frequencies change due to changes in $(\Gamma_1 p)$ 
and $\rho$.
   This is described and discussed in Appendix~\ref{app:a}.

\begin{figure}
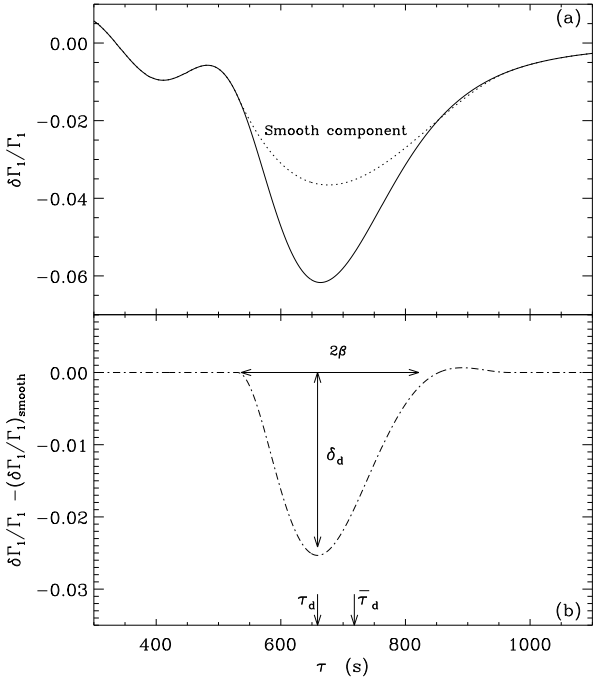

\centering
\infig{fig02}
 \caption{(a) Plot of the differences $(\delta\Gamma_1/\Gamma_1)$ between
two models - one with and the other without the second ionization of
helium, versus acoustic depth $\tau$.
  These correspond to models $Z_0$ and $Z_1$ discussed in the text and
described in Table~1.
  The dotted line represents our assumed smooth reference structure.
  (b) The change of $\Gamma_1$, relative to the smooth reference
structure, is shown.
  The parameters represented schematically, to be determined from the
characteristics of the signal in the frequencies, are: $\deltad$ - the
amplitude of the differences at $\taud$, and $\beta$ - the half width of
$\delta\Gamma_1/\Gamma_1$ (values taken from Table~\ref{tab:par}, below).
  Also indicated is the value of $\btaud$ as found from the frequencies.}
 \label{fig:delta-gamma1}
\end{figure}

  In order to model the signature of the ionization zone we represent the
effect of the second ionization in terms of the changes it induces in the
adiabatic exponent $\Gamma_1$.
   Such a change (see Fig~\ref{fig:delta-gamma1}) is approximately
represented by a `bump' of half width $\beta$ in acoustic depth, and
relative height
 \begin{equation} 
\deltad \equiv \left(\dgamma \right)_{\taud} ,
 \end{equation}
being located at a radial position corresponding to an acoustic depth $\taud$.
   Here, and in the following, acoustic depth $\tau$ at a radius $r$
is defined as,
\begin{equation}
  \tau(r) \equiv \int_r^R {\diff r \over c} \;,
 \label{eq:tau}
\end{equation}
where $R$ is the photospheric radius of the Sun. 

  Relatively to the frequencies of a reference model, assumed to be
`smooth' and corresponding approximately to a model with no \HeII
ionization region, we find that the bump changes the frequencies in such a
way that there is a periodic component of the form (see
Appendix~\ref{app:a}),
 \begin{equation} 
 \delta\omega \sim A(\omega,l)\; \cos\Lambdad \;,
 \label{eq:signal}
 \end{equation}
 where the amplitude, as a function of mode frequency $\omega$ and mode
degree $l$, is given by
 \begin{equation}
A(\omega,l) \equiv a_0\;
      {1{-}2\del/3 \over \left(1{-}\del\right)^2}\;\;
      {\sin^2\left[\beta\omega\left(1{-}\del\right)^{1/2}\right]
      \over \beta \omega}\;,
\label{eq:signal_amp}
 \end{equation}
and the argument is
 \begin{equation}
\Lambdad \equiv 2 \left[
   \omega \int_0^{\taud} \left(1{-}\del\right)^{1/2}\diff\tau
   + \phi \right] \;.
 \end{equation}
  Here the factor in $\Delta$ represents the geometry of the ray-path
accounting for deviation from the vertical when the mode degree is non-zero.
  It is associated with the Lamb frequency, as given below 
(Eqs~\ref{eq:del}, \ref{eq:deld}). 
  In fact, because the ionization zone is close to the surface and 
provided we are not using very high-degree data, we can 
neglect $\del$ in the expression for the argument $\Lambdad$; we can similarly 
neglect the effect of the 
mode degree on the surface phase function $\phi$.
  Consequently, for the ionization zone the expression of the argument
becames
 \begin{equation}
\Lambdad  \sim 2\, ( \omega \taud + \phi)
   \simeq 2 \left(\omega\btaud + \phi_0 \right) \;.
 \label{eq:lambdad}
 \end{equation}
  In the asymptotic expression used for the eigenfunction
(see Eq.~\ref{eqa:eigen}), the phase $\phi$ depends on the
mode frequency and degree (see MCDT for details).
  Here we have expanded the phase to first order
in frequency, by writing that $\phi(\omega){\simeq}\phi_0{+}a_\phi \omega$.
  From this it follows that $\btaud{\equiv}\taud{+}a_\phi$, while
the frequency independent term of the phase is now $\phi_0$.

  The amplitude of the signal, through $a_0$, corresponds to
 \begin{equation}
a_0 = {3 \deltad \over 2\taut} \;,
 \end{equation}
  where $\taut{\equiv}\tau(0)$ is the total acoustic size of the Sun.
  The small factor $\del$, present in the amplitude, is given by
 \begin{equation}
\del = \deld\; {l(l{+}1) \over \ltil(\ltil{+}1)}\;
            {\omegat^2 \over \omega^2} \;,
\label{eq:del}
 \end{equation}
  where the value of $\deld$ is given by,
 \begin{equation}
\deld = {\ltil(\ltil{+}1) \over \omegat^2} \,
   \left({c \over r}\right)^2_{\tau{=}\taud} \;,
\label{eq:deld}
 \end{equation}
  and $\ltil$ and $\omegat$ are two reference values.
  These values are chosen taking into account the expected behaviour of
the signal and the set of modes used, as discussed below.

   In order to compare the amplitude it is convenient to define a
reference value $\Ad$, as given by
 \begin{equation}
\Ad \equiv  A(\omegat,\ltil) = a_0\;
      {1{-}2\deld/3 \over \left(1{-}\deld\right)^2}\;\;
      {\sin^2\left[\beta\omegat\left(1{-}\deld\right)^{1/2}\right]
      \over \beta \omegat} \;.
\label{eq:amp_ref}
\end{equation}

  The parameters of the signal relevant to characterize the local
properties of the ionization zone, as given in Eq.~(\ref{eq:signal}), are
$\btaud$, $\beta$, $a_0$ and $\deld$.

  The values of $\btaud$ and $\deld$ can be used to measure mainly
the location of the ionization zone.
  They both vary strongly with distance to the surface.
  But the acoustic depth is a cumulative function of the sound
speed behaviour over all layers from the surface to a particular position,
  whereas $\deld$ is a local quantity, not being affected
by the behaviour of the sound speed in the layers above
the ionzation zone.

  The values of $\beta$ and $a_0$ (or $\deltad$) are
expected to be directly related to the local helium abundance,
since the size of the bump will be determined by the amount of
helium available to be ionized.
  These parameters are also expected to be strongly affected by
the equation of state, and to a lesser extent by the other
physics that affect the location of the ionization zone ($\taud$).
   But we may hope to be able to use the other parameters to remove this
dependence, while retaining the strong relation between the bump and the
helium abundance ($Y$).

\subsection{Measuring the signal in the frequencies}\label{sec:signal}

  Our first goal is to find the five parameters describing the signal from the
frequencies of oscillation.
  In order to do that we must isolate a signature of about 1$\muHz$
in amplitude, overimposed in actual frequencies.
  We do so by isolating in the frequencies the periodic signal described
by Eq.~(\ref{eq:signal}) using a non-linear least-squares iterative fit to
find the best set of parameters.
  The method used is an adaptation of the one proposed by MCDT; for
the present problem we must redefine the characteristic wavelength of the
signal to be isolated (quantity $\lambda_0$ in MCDT) as it is significantly
larger than for the signal from the base of the convective envelope.
   The parameters describing the signal (Eq.~\ref{eq:signal}), and found
by our fitting procedure, are the following;
$$
 \taud,\quad \phi_0,\quad a_0,\quad \deld,\quad \beta \;.
$$
   We choose a set of modes which cross the
ionization zone, but which do not cross the base of the convection zone.
   By removing modes that penetrate deep in the Sun (low degree modes),
we avoid the contamination coming from the signal generated at the 
base of the convection zone (see MCDT).
But when selecting only modes of higher degree (between 45 and 100),
it becomes necessary to include the contribution from the mode degree
to the amplitude of the signal.
This is the reason why it is necessary to include in the fitting
the parameter $\deld$.
  This parameter is not necessary when studying other stars
\citep{monteiro98,basu04,piau05}, resulting in a
simplified description of the expected observed behaviour.
In the case of the Sun there is a great advantage in using all
available high-degree modes that cross the ionization zone.

   The modes considered correspond to the ones available in solar data,
having degrees and frequencies such that the lower turning point is
between $0.75R$ and $0.95R$ of the solar radius.
   The latter ensures the modes cross the ionization zone while the former
avoids contamination from the signal originating at the base of the
convective envelope (\eg\ CDMT, and references therein).
   These conditions define typically a set of about 450 modes having
frequency $\omega/2\pi$ in the range $[1500,3700]~\muHz$, and with mode
degree of $45{\le}l{\le}100$.

   As we are only using modes of high degree in this work,
the reference values preferred in the fitting of the signal are;
$$
\ltil = 100  \quad{\rm and}\quad 
{\omegat \over 2\pi} = 2000\muHz \;.
$$
   The first value is an upper limit for modes that cross beyond the
ionization zone, while the value of $\omegat$ corresponds to the region in
frequency where the signal is better defined.
   These values are only relevant to normalize the parameters
fitted for different models.

   For solar observations only frequencies with a quoted observational
error below 0.5 $\muHz$ are included.
   We ensure consistency of the data sets by restricting the
selection of mode frequencies from the models to the modes
present in the solar data after applying the above selection rules.

   We stress that the method adopted for removing the
smooth component of the frequencies is a key assumption in the
process of fitting the signal.
   Here we use a polynomial fit with a smoothing parameter on the third
derivative (see CDMT).
   In any case, as long as the analyses for different models and for the
solar data are consistent, the comparison of the parameters
will be independent of the choice on how to describe the smooth component.
   Such consistency is ensured by using exactly the same set of
frequencies and the same numerical parameters for the fitting.

\subsection{The signal in the solar data}

\begin{figure}
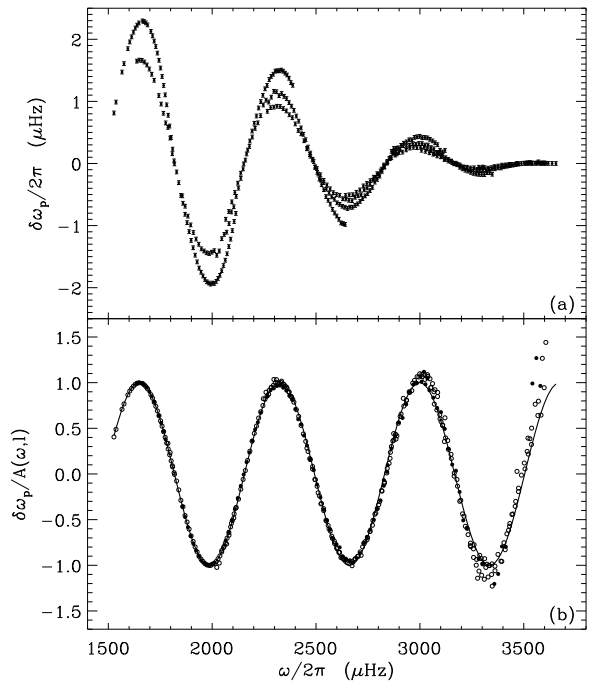
 
\centering \infig{fig03}
 \caption{(a) Residuals left after a smooth component of the frequencies,
as functions of mode order and degree, is removed.
  The data are from MDI/SoHO, with the error bars corresponding to a
3$\sigma$ of the quoted observational errors.
  (b) Plot of the signal isolated, and shown in panel (a), after division
by the amplitude function as given by Eq.~(\ref{eq:signal_amp}) when using
the values of $(a_0,\deltad,\beta,\deld)$ found in the fitting.
  Modes with degree below $l{=}60$ are shown as filled circles, while
modes with a higher value of the degree are represented by open circles.
  The line indicates the fitted periodic signal as expected from
Eq.~(\ref{eq:signal}).}
 \label{fig:signal_sun}
\end{figure}

   To illustrate the signal extraction, the method
proposed here was applied to the analysis of solar seismic data from
MDI on the SOHO spacecraft \citep{scherrer95}.
   The signal was isolated as described above for the models.
   The fitted signal of the Sun is shown in Fig.~\ref{fig:signal_sun}a,
together with the error bars.
   In order to illustrate how well the expression for the signal
(Eq.~\ref{eq:signal}) fits the data points we also show in
Fig.~\ref{fig:signal_sun}b the signal in the frequencies normalized by the
amplitude as given in Eq.~(\ref{eq:signal_amp}).
  The quality of the fit done with Eq.~(\ref{eq:signal}), confirms the
adequacy of the first order analysis developed in Appendix~\ref{app:a}
leading to the expression given by Eq.~(\ref{eq:signal_amp}).

\begin{table}
  \begin{center}
    \caption{Parameters obtained by fitting the observed solar
frequency data with the expression of the signal as given in
Eq.~(\ref{eq:signal}).
  The quantities $\taud$ and $\beta$ are given in seconds, while the amplitudes
($a_0$ and $\Ad$) are given in $\muHz$.
Note that $\Ad$ is not a fitting parameter, as it is given from the other
parameters using Eq.~(\ref{eq:amp_ref}).
The standard deviations $\sigma$ are estimated from 
200 simulations of the effect
of the observational uncertainties on the determination of the parameters.}
    \begin{tabular}[h]{crccrrc}
      \hline
       & $\bar\tau_d\;$ & $\phi_0$ & $a_0/2\pi$ & $\Ad/2\pi$ &
              $\beta\;\;$ & $\deld\;$ \\[+2pt]
      \hline
 Sun     & 741.2 & 1.743 & 1.987 & 2.655 & 141.3 & 0.493 \\[5pt]
 $3\sigma$&  1.9 & 0.027 & 0.056 & 0.037 & 0.96& 0.015 \\
      \hline
      \end{tabular}
    \label{tab:par_sun}
  \end{center}
\end{table}

   The values of the parameters found for the data are given in
Table~\ref{tab:par_sun}.
   From Monte Carlo simulations we have estimated the uncertainty in the
determination of the parameters due to observational uncertainties as
indicated by the quoted observational errors.
   The values found, at the 3$\sigma$ level, are also listed in
Table~\ref{tab:par_sun}.
   It is clear that due to the large amplitude of this signature
(above $1\muHz$ at $\omega/2\pi{=}2000\muHz$) the
precision with which the parameters are determined is very high.
   As long as the method to isolate this characteristic signature is
able to remove the ``smooth'' component, the results can be used
with great advantage for testing the zone of the
second ionization of helium in the Sun.

\section{Solar models with different physics}\label{sec:models}

  In order to establish how sensitive the different characteristics of
the signal are to the properties of the ionization zone, and therefore to the
aspects of the Sun which affect the ionization zone, we consider
different static models of the Sun calculated with the
same mass, photospheric radius and luminosity.
   The profile of the helium abundance in the models is obtained by
calibrating with a constant factor a prescribed abundance profile from
an evolution model with the age of the Sun (without settling).

   We note that imposing the same radius and luminosity for all models
is the key difference between the analysis presented here
and the work by \cite{basu04}.
  If the models are not required to have the same luminosity and
radius as the Sun, the properties of the ionization zone are not affected
in the same way. 
   Consequently the behaviour of the amplitude of the signal 
in this case is different
from what we find when both these conditions are imposed on the models.

   The aspects of the physics being tested here are
the {\it equation of state (EoS)},
the {\it theory of convection} and 
the {\it opacity}.
   All these aspects affect the ionization zone by changing its location,
size and thermodynamic properties.

\begin{table}
  \begin{center}
    \caption{Solar models and their helium ($Y$) abundances.
 Also indicated are the equation of state (EoS):
 SEoS - Simple Saha equation of state with pressure ionization, and
 CEFF (cf. \citealt{cd92a});
 the Opacity:
 SOp - simple power law fit of the opacities, and
 Kur - low temperature opacities from \citet{kurucz91};
 and the formulation for modelling convection:
 MLT - standard mixing length theory (\citealt{vitense58}; parametrized
 as in \citealt{monteiro96}), and
 CGM - \citet*{canuto96}.
 See the text for a description of the parameter $f_\epsilon$ used in
 the calculation of the emissivity.}
    \begin{tabular}[h]{ccccccccc}
      \hline
        Model & EoS & Opacity & Convection & $Y$ & $f_\epsilon$ \\
        \hline
        $Z_0$    & SEoS & SOp & MLT & 0.24615 \\[+5pt]
        $Z_1$    & SEoS & SOp & MLT & 0.24608 \\[+5pt]
        $Z_2$    & SEoS & SOp & CGM & 0.24608 \\[+5pt]
        $Z_{3l}$ & CEFF & SOp & MLT & 0.24149 & 0.8 \\
        $Z_3$    & CEFF & SOp & MLT & 0.24981 \\
        $Z_{3h}$ & CEFF & SOp & MLT & 0.25667 & 1.2 \\[+5pt]
        $Z_4$    & CEFF & SOp & CGM & 0.24981 \\[+5pt]
        $Z_{5l}$ & CEFF & Kur & MLT & 0.24148 & 0.8 \\
        $Z_5$    & CEFF & Kur & MLT & 0.24980 \\
        $Z_{5h}$ & CEFF & Kur & MLT & 0.25667 & 1.2 \\
        $Z_{5v}$ & CEFF & Kur & MLT & 0.26246 & 1.4 \\[+5pt]
        $Z_6$    & CEFF & Kur & CGM & 0.24980 \\
        \hline
      \end{tabular}
    \label{tab:models}
  \end{center}
\end{table}

\begin{figure}
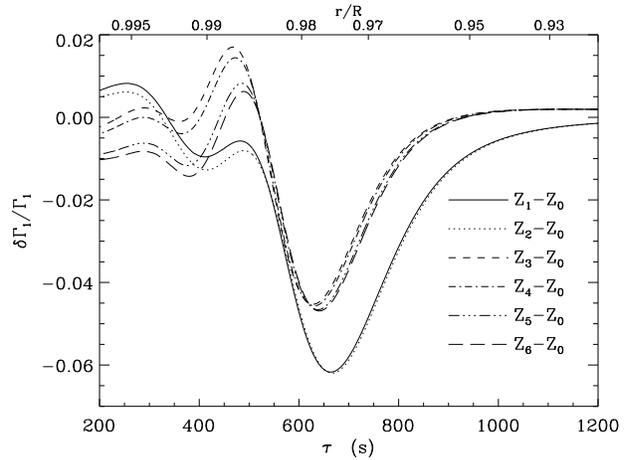

\centering
\infig{fig04}
 \caption{Plot of the differences for $\Gamma_1$ between all models
considered and the one without the second ionization of helium.
  See Table~\ref{tab:models} for the details of each model.
  Only the region around the second ionization of helium is shown
corresponding to the negative bump around an acoustic depth of 650s. }
 \label{fig:dif-gamma1}
\end{figure}

  All models were calculated as in Monteiro (\citeyear{monteiro96t};
see also \citealt{monteiro96}).
  These are not intended to represent accurately the Sun, but simply to
illustrate the usefulness of the method we propose to study a particular
region of the solar envelope.

  As the simplest possible EoS we have used a Saha equation of state with
full ionization at high pressure - this corresponds to SEoS in
Table~\ref{tab:models}.
   As a more complete EoS we have used the CEFF equation of state as
described in \citet{cd92a}.
   For the opacities we have considered a simple power law fit (SOp), or the
Rosseland mean opacity tables at low temperatures from \citet{kurucz91}.
   To include convection we have taken the standard mixing length theory
(\citealt{vitense58}, parametrized as in \citealt{monteiro96}) or the more
recent CGM model \citep{canuto96}.

   As our reference model, in order to illustrate the changes due to the
ionization of helium, we have calculated a very simple solar model ($Z_0$)
with suppressed \HeII ionization, by setting the ionization potential to
zero.
   The helium abundance found for each model corresponds to the value that
fits the boundary conditions (by scaling a prescribed dependence of the
chemical profile, as taken from an evolved solar model).

\begin{figure}
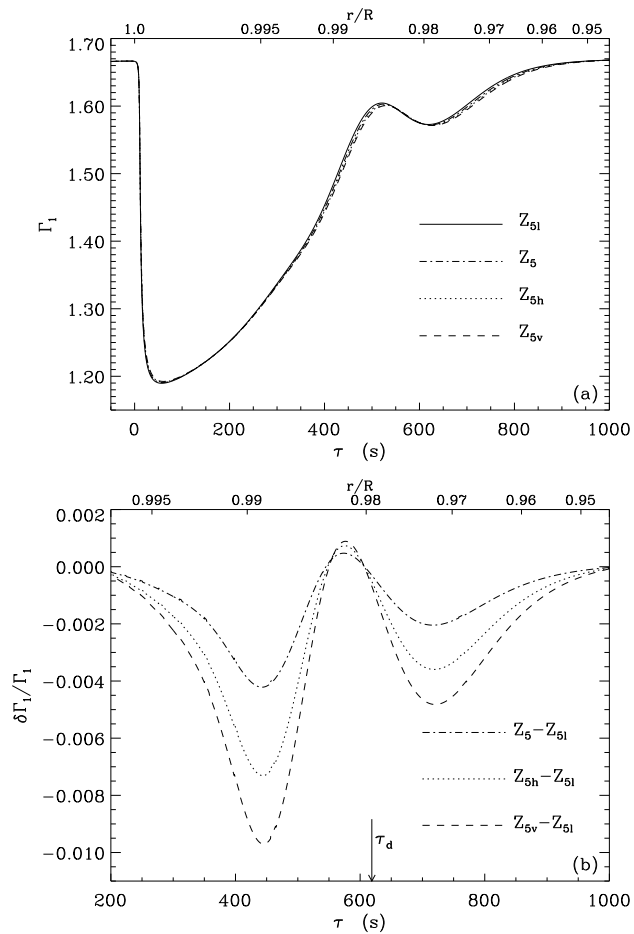

\centering
\infig{fig05}
 \caption{(a) Plot of the adiabatic exponent $\Gamma_1$ for all models `5',
calculated with the same physics but different surface helium abundances
as imposed by $f_\epsilon$ (see Table~\ref{tab:models}).
The hydrogen and both helium ionization regions are shown.
(b) Plot of the differences for $\Gamma_1$ between models $Z_{5,5h,5v}$
and model $Z_{5l}$.
  See Table~\ref{tab:models} for the details of each model.
  Only the region around the second ionization of helium 
  (indicated by the arrow) is shown. }
 \label{fig:dif5-gamma1}
\end{figure}

  The behaviour of the adiabatic exponent for some of the models (see
Table~\ref{tab:models}),
relative to our reference model ($Z_0$), is
illustrated in Fig.~\ref{fig:dif-gamma1}.
   There is a clear difference on the location of the ionization zone
($\taud$) when a different EoS is used.
   The effects of changes in the formulation of convection or in the
opacities are much smaller.

  In order to have models with the same envelope physics, but different
helium abundances, we have calculated solar models with the energy
generation rate changed by a prescribed factor $f_\epsilon$ in the emissivity.
   These are models $Z_{3l,3h}$ and $Z_{5l,5h,5v}$ which are similar to
$Z_3$ and $Z_5$, respectively, except for the value of $f_\epsilon$ which
is now different from unity.
  These correspond to models with a different structure of the core but with
envelopes with exactly the same set of physics.
All differences between these models in the envelope are due to differences
in the chemical composition. 
  To illustrate the differences we plot in
Fig.~\ref{fig:dif5-gamma1} the differences in $\Gamma_1$ between models
with the same physics but increasing values for the envelope abundance of
helium.
  As the helium abundance increases, there is a corresponding decrease in
hydrogen, which results on a slight separation in temperature of the
three major ionization regions.
  Consequently both ionization regions for the helium expand towards
higher temperatures.
  As the bump becomes slightly wider and moves to a higher temperature,
the effect on the frequencies is
expected to become smaller.

  For all models we have calculated the frequencies of linear adiabatic
oscillations.
  The set of frequencies for each model, as used to fit the signature
of the ionization zone, is described above.
  The parameters obtained in fitting Eq.~(\ref{eq:signal}) to the frequencies
of the models listed in Table~\ref{tab:models} (excluding $Z_0$) are given in
Table~\ref{tab:par}.

\begin{table}
  \begin{center}
    \caption{Parameters obtained by fitting the frequency data for the
models with the expression of the signal as given in
Eq.~(\ref{eq:signal}).
  The quantities $\taud$ and $\beta$ are given in seconds, while the amplitudes
($a_0$ and $\Ad$) are given in $\muHz$.
Note that $\Ad$ is not a fitting parameter, as it is given from the other
parameters using Eq.~(\ref{eq:amp_ref}).}
    \begin{tabular}[h]{crccrcc}
      \hline
      Model & $\bar\tau_d\;$ & $\phi_0$ & $a_0/2\pi$ & $\Ad/2\pi$ &
              $\beta$ & $\deld\;$ \\[+2pt]
      \hline
 $Z_1$   & 718.0 & 2.588 & 1.634 & 2.834 & 142.5 & 0.604 \\[+5pt]
 $Z_2$   & 724.8 & 2.525 & 1.671 & 2.862 & 141.9 & 0.599 \\[+5pt]

 $Z_{3l}$& 729.9 & 1.950 & 2.500 & 3.251 & 146.0 & 0.484 \\
 $Z_{3}$ & 730.4 & 1.951 & 2.380 & 3.140 & 144.7 & 0.490 \\
 $Z_{3h}$& 730.4 & 1.951 & 2.314 & 3.066 & 144.3 & 0.491 \\[+5pt]

 $Z_4$   & 739.9 & 1.859 & 2.353 & 3.151 & 143.3 & 0.495 \\[+5pt]

 $Z_{5l}$& 737.7 & 1.874 & 2.429 & 3.241 & 143.7 & 0.494 \\
 $Z_{5}$ & 737.8 & 1.876 & 2.342 & 3.145 & 143.1 & 0.496 \\
 $Z_{5h}$& 737.5 & 1.880 & 2.278 & 3.072 & 142.7 & 0.498 \\
 $Z_{5v}$& 736.8 & 1.890 & 2.205 & 3.002 & 141.7 & 0.502 \\[+5pt]

 $Z_6$   & 746.4 & 1.790 & 2.280 & 3.141 & 141.3 & 0.507 \\
      \hline
      \end{tabular}
    \label{tab:par}
  \end{center}
\end{table}

\section{The effect of the physics on the characteristics
         of the signal}\label{sec:results}

  The set of solar models considered here, and listed in
Table~\ref{tab:models}, cover three major aspects of the physics which
determine the surface structure of the models:
  the equation of state, the low temperature opacities and the
formulation for convection (defining the superadiabatic layer).
  In order to use the diagnostic potential of this characteristic
signature in the frequencies we need to understand how each parameter
describing the signal is affected by a specific aspect of the physics
defining the structure of the envelope.

  One would expect that the shape of the bump is strongly determined by
the EoS.
  But any change in the structure that affects the location of the
ionization zone will necessarily introduce an effect on the
characteristics of the $\Gamma_1$ profile.
  Consequently we need first to identify the parameters that depend more
strongly on the location.
  These are most likely $\btaud$ and $\deld$.

  The changes on the upper structure of the envelope are expected to have
a direct effect on the turning point of the modes.
  Consequently we need to look at the parameters that may be affected by
the upper reflecting boundary.
  This is mainly expected to be $\phi_0$.

  Finally the area of the bump in $\Gamma_1$ in the ionization zone
should reflect the local abundance of helium,
if the location is well defined.
  Therefore we will look at $a_0$ and $\beta$ in order to identify how 
the helium abundance $Y$ defines the
characteristics of the signal in the frequencies.

\subsection{The location of the ionization zone}\label{sec:res_loc}

  The most easily identifiable characteristic of the signal is its period.
  This quantity depends strongly on $\taud$, but as discussed when writing
Eq.~(\ref{eq:lambdad}) the period also contains a 
contribution from the upper turning
point of the modes (where there is a phase shift of the eigenfunction).
  This means that the period, or precisely $\btaud$, that we measure 
is not necessarily a good estimate of location $\taud$ 
of the ionization zone.

\begin{figure}
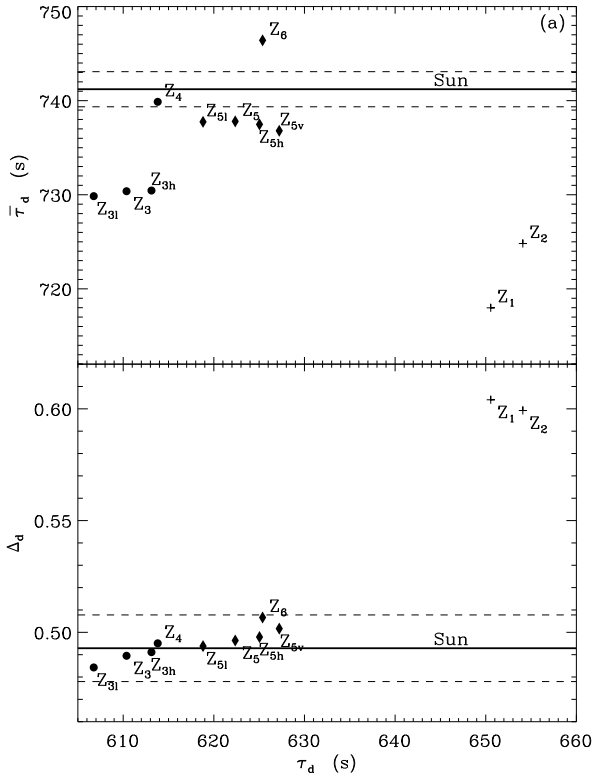

\centering
\infig{fig06}
 \caption{(a) Plot of the fitted acoustic depth $\btaud$ and (b)
the correction term $\deld$, versus the acoustic depth $\taud$ as determined
from the models and corresponding to the local minimum in $\Gamma_1$ (see
Fig.~\ref{fig:gamma1}).
  Filled symbols are for models using the CEFF equation of state, while
crosses are for models calculated using a simple Saha equation of state.
  The filled circles are for models having the same simple opacity (power
law) but different theories of convection, while the filled diamonds are
models with the opacity at low temperatures from Kurucz.
  The values found for the solar data are also shown in both panels with
$3\sigma$ error bars (dashed horizontal lines) due to the observational
uncertainties.}
 \label{fig:btaud_deld}
\end{figure}

  Figure~\ref{fig:btaud_deld}a shows the value of $\btaud$, as found
from fitting the signal in the frequencies, versus the value of $\taud$,
as determined from the location of the local minimum of $\Gamma_1$ in
the model.
  There is a difference of up to about 140~s between $\btaud$ and
$\taud$, and one is not simply a function of the other.
  The difference between the two comes from $a_\phi$, which measures the
leading-order frequency dependence of the phase transition which the
eigenfunctions undergo at the upper turning point.
  This will be strongly affected by the physics that change the
structure of the surface, namely convection, EoS, the low temperature
opacities, and the structure of the atmosphere.
  Consequently, we have to use some caution when taking the parameter
$\btaud$ from the fit to estimate the location of the ionization region.

  As an alternative we can consider one of the other parameters which
also depends on the position of the ionization zone.
  This is $\deld$, given in Table~\ref{tab:par} for all models and shown
in Fig.~\ref{fig:btaud_deld}b as a function of the actual acoustic
location of the ionization region.
  The value of $\deld$, defined in Eq.~(\ref{eq:deld}), is not sensitive
to the layers near the photosphere, as its value is determined exclusively by
the sound speed at the ionization zone.
   However, the determination of this term is associated with a small
correction in the amplitude, which makes it more sensitive to the
observational errors when fitting the frequencies.

  Both panels in Fig.~\ref{fig:btaud_deld} show the solar values of
$\btaud$ and $\deld$ with 3$\sigma$ uncertainties.
  The values of $\deld$ indicate that all models calculated with the
CEFF equation of state give, even if marginally, a location for the
ionization zone consistent with the Sun.

\begin{figure}
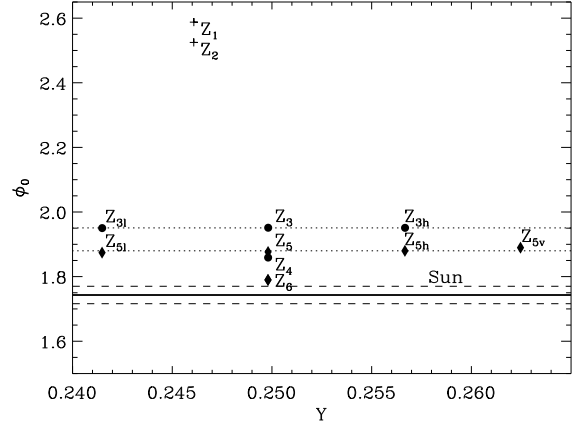
 
\centering \infig{fig07}
 \caption{Plot of the phase $\phi_0$ of the signal versus the envelope
helium abundance $Y$ for all models.
  The symbols are the same as in Fig.~\ref{fig:btaud_deld}.
  The dotted lines illustrate the correlation between models with the same
physics but different values of the surface helium abundance $Y$.
  The value found for the solar data is also shown with $3\sigma$ error
bars (dashed horizontal lines) due to the observational uncertainties.}
 \label{fig:par_phase}
\end{figure}

  Finally, the structure at the top of the envelope is also expected to be
reflected in the value of $\phi_0$.
  The value of this parameter for all models is represented in
Fig.~\ref{fig:par_phase} as a function of the envelope helium abundance.
  The larger difference is found when changing the EoS (about
$0.06$).
  But changes in the opacities also change $\phi_0$ by as much as
$0.01$, while the theory of convection changes this by about
$0.01$.
  It is interesting to confirm that the fitted value of $\phi_0$ is
independent of the helium abundance, as one would expect from the analysis
leading to the expression of the signal.
  Consequently $\phi_0$ may allow the separation between the helium
abundance and the physics relevant to the outer layers of the Sun 
because it is insensitive to $Y$ whilst being indicative of some
near-surface change that may be required in the physics.

  The solar value for $\phi_0$ is also shown in Fig.~\ref{fig:par_phase}.
  Adjustments in the near surface layers seem to be necessary in order
to produce models that have a value of $\phi_0$ consistent with the Sun.
  Changes in the superadiabatic layer or in the surface opacities may
be some of the options for reconciling the models with the solar data.

\subsection{The equation of state}\label{sec:res_eos}

\begin{figure}
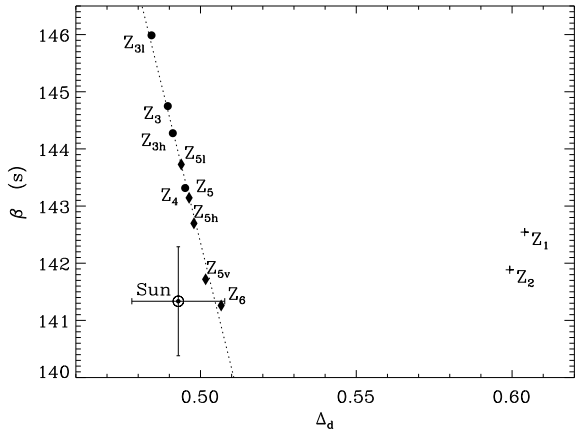

\centering
\infig{fig08}
 \caption{Plot of the estimated width of the bump, as given by $\beta$,
versus the value $\deld$ providing an indication of the location of the
ionization zone.
  The symbols are the same as in Fig.~\ref{fig:btaud_deld}.
  The dashed line indicates a linear fit to the models with the same EoS.
  The values found for the solar data are also shown with $3\sigma$ error
bars due to the observational uncertainties.}
 \label{fig:beta_deld}
\end{figure}

  From the analysis of the results listed in Table~\ref{tab:par}, and as
discussed in the previous section, the EoS is the most important factor in
defining the characteristics of the signal.
  In Fig.~\ref{fig:beta_deld} we show the width parameter $\beta$
as a function of $\deld$ (a proxy for the location).
  Models that have the same EoS (CEFF) lie on a common locus in this diagram, 
as indicated by the dotted line.
  The position along this line of models all built with the CEFF
varies according to changes in the convection or the surface
opacities.
  Models $Z_1$ and $Z_2$, built with a different EoS, lie
in a different region of the diagram.
  Thus we claim that, with the location of the ionization zone fixed,
the width of the bump in $\Gamma_1$ is mainly a function of the EoS,
as expected. 

  Consequently, after using $\phi_0$ to test the surface physics, it is
possible to combine the constraints provided by $\deld$ and
$\beta$ to obtain a direct test on the EoS and
the location of the ionization zone.

  Figure~\ref{fig:beta_deld} also includes the parameters found for the
solar data.
  These are marginally consistent with the expected behaviour found using
models calculated with the CEFF equation of state.
  Other options for the EoS must be considered in an attempt to bring the models
closer to the Sun.

\subsection{The helium abundance in the envelope}\label{sec:res_helium}

  From the discussion in the previous sections it follows that any
determination of the helium abundance requires a careful tuning of the
models to the correct structure of the envelope.
   Such a fine tuning can be performed based on the sensitivity of the
eigenfrequencies to the behaviour of the adiabatic exponent in the region
where helium undergoes its second ionization.
   We have found, as discussed above, that:
 \begin{itemize}
 \item $\deld$ provides a process to place the ionization zone in the
model at the same acoustic depth as for the Sun --
this corresponds to adapting mainly the surface layers of the model
(atmosphere and/or convection) in order to place the ionization
zones at the same acoustic location as measured in the Sun by the
solar value of $\deld$;
 \item $\beta$ can then be used to adjust the EoS (or more likely to 
select it from a few candidates) to match the observed
behaviour -- this corresponds to verifying that the behaviour of $\beta$
as a function of the location ($\deld$) in the models includes the
observed solar values for these two parameters;
 \item and finally, the parameters $\btaud$ and $\phi_0$ can be combined
to adjust the surface physics in the model, in order to
recover the observed solar values -- this corresponds to adjusting
convection (superadiabatic region, mainly), opacities
(low temperature range), photosphere, etc, in a complementary way
to the first point, until the solar values can be recovered with
the models as both parameters are strongly dependent on these
aspects of the physics, but quite insensitive to the actual helium abundance.
 \end{itemize}

  Consequently, we are left with one last parameter, connected with the
shape of the bump through $\deltad$, which is the amplitude of the signal
$a_0$, or $\Ad$.
  If the model has been adjusted to the observed data using the remaining
parameters, then the amplitude will depend mainly on the helium abundance
in the model, which can now be compared with the solar abundance.
  Such a relation provides a 
measurement of the helium abundance,
which complements the boundary condition used in the evolution to fit the
model to the present day Sun.

\begin{figure}
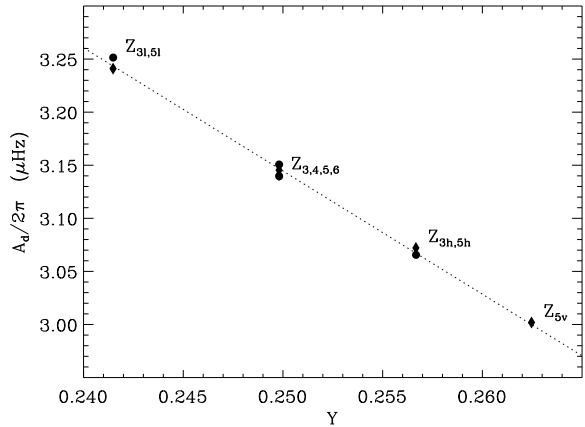

\centering
\infig{fig09}
 \caption{Plot of the reference amplitude (see Eq.~\ref{eq:amp_ref})
versus the envelope helium abundance $Y$.
  The symbols are the same as in Fig.~\ref{fig:btaud_deld}.
  The dotted line illustrates the correlation between the helium abundance
$Y$ and the amplitude.}
 \label{fig:sig_amp}
\end{figure}

  Figure~\ref{fig:sig_amp} illustrates how such a dependence of $\Ad$, as
defined in Eq.~(\ref{eq:amp_ref}), could be constructed after the other
aspects of the physics are adjusted.
  It is worth noting that, as expected from Fig.~\ref{fig:dif5-gamma1},
the amplitude {\it decreases} with increasing $Y$, since the changes in
$\Gamma_1$ due to changes on the hydrogen abundance dominate the
variations of the bump.
  This regime for the inverse dependence of the amplitude of the signal
on the abundance of helium is relevant for stars of low effective temperature.
  That follows from the overlapping of the three ionization zones
(H{\sc i}, He{\sc i}, \HeII).
  For stars where these are fully separated in temperature it is
expected that the amplitude will {\it increase} with the abundance of helium.

  As shown above (see Figs~\ref{fig:par_phase} and \ref{fig:beta_deld}) the
models used here are not fully consistent with the physics of the Sun
and seem to be only marginally consistent regarding the equation of state
that has been used.
  Consequently the amplitude $\Ad$, as found for the solar data, cannot yet 
be used as an indicator of the helium abundance in the solar envelope.
  A more adequate calibration of the surface layers in the models must be
developed before an estimation for $Y$ is inferred from this parameter.

  The simplified models we are using here to illustrate
the applicability of the method have been calculated with scaling a chemical
profile determined without including diffusion and settling of helium.
  This is one of the aspects that needs to be considered in the models in
order to move the parameters found for these closer to the solar values.
  With such a tuning, based on other seismic constraints and on
the parameters of the signal discussed here, we have
an independent procedure to adjust our models to the Sun in this
region near the surface, where the uncertainties in the physics dominate
the structure of the models.

\section{Conclusion}\label{sec:dis}

  In this work we have developed a complementary method to constrain the
properties of the helium second ionization region near the surface of the
Sun using high degree mode frequencies.
  The method proposed here can independently test properties of this
region, and provides a possible direct measurement of the helium abundance
in the envelope.

  We have shown that some of the parameters characterizing the signature
in the frequencies due to this region in the Sun are very sensitive to the EoS
used in the calculation of the models, and so can be used to test and
constrain the equation of state.
   Others of the parameters can also provide an important test on the
physics affecting the surface regions of the models, namely convection and
the low temperature opacities.
   By combining the diagnostic potential of the five parameters
determined from the data with very high precision the helium abundance
can be effectively constrained.

  Here we were mainly concerned to establish the method and demonstrate
how it can be used to study the \HeII ionization zone in the Sun, and the
physics that affect the structure of the Sun in that region.
   In spite of having used simplified models to represent the
Sun we have illustrate the sensitivity of each parameter to the physics,
establishing the approach that can be followed when adequate up-to-date
evolved solar models are used.
   Besides the physical ingredients addressed here, aspects like
diffusion and settling and improved opacities have to be implemented
in order to provide a physically consistent value of the helium abundance.
A calibration of the actual solar helium abundance using models with the
best up-to-date physics will be the subject of the second paper in this
series.

\section*{Acknowledgements}

  We are grateful to S. Basu, J. Christensen-Dalsgaard, M.P. di Mauro
and  A. Miglio for valuable discussions.
  This work was supported in part by the Portuguese {\it Funda\c{c}\~ao
para a Ci\^encia e a Tecnologia} through projects
{\scriptsize POCTI/FNU/43658/2001} and {\scriptsize POCTI/CTE-AST/57610/2004}
from POCTI, with funds from the European programme FEDER.


\appendix
\section{A variational principle for the \HeII ionization
         zone}\label{app:a}

   We consider here a variational principle, following the procedure
by \cite{monteiro96t}, for describing how the modes
are affected by the presence of the region of the second ionization of the
helium.
   We start by using a variational principle, for small changes of the
eigenfrequencies ($\omega$) due to small changes of the structure.
It can be written (see \citealt{cd95},
and references therein) in the form
 \begin{equation}
\delta\omega^2 = {\delta I\over I_1}
\qquad {\rm with}\qquad
I_1 \sim {1\over 2}\; \taut E_o^2 \;.
\label{eqa:var_princ}
 \end{equation}
  Here, $\taut$ is the acoustic size of the Sun, and
 \begin{equation}
\delta I {\sim} \!\int_0^{\tau_t}\!\! {\left[ \left( \delta B_1 {+}
        \devdf{\delta B_0}{\tau}\right) E_r^2 {+}
        \delta B_2\; \devdf{E_r^2}{\tau} {+}
        \delta B_3\; \devds{\!E_r^2}{\tau} \right] \diff\tau} \,,
\label{eqa:deltai}
 \end{equation}
  where $E_r$ is the normalized radial component of the eigenfunction (with
constant amplitude $E_0$).
  The acoustic depth $\tau$ is defined in Eq.~(\ref{eq:tau}).

  From asymptotic analysis (see MCDT) we also have that well inside the
turning points and for moderate degree modes,
 \begin{equation}
\devds{E_r}{\tau} \sim -\omega^2\left(1{-}\del\right)\; E_r \;,
 \end{equation}
  or
 \begin{equation}
E_r \sim E_0 \cos\left[\omega\int_0^\tau
    \left(1{-}\del\right)^{1/2}\diff\tau + \phi \right] \;.
\label{eqa:eigen}
 \end{equation}
  The changes in the structure relative to the reference (`smooth')
model are described with the functions $\delta B_i$, as given by
 \begin{equation}
{\delta B_0 \over g/c} = - \drho \;,
\label{eqa:b0}
 \end{equation}
 \begin{eqnarray}
{\delta B_1 \over \omega^2} &=& \bigg\{
        - {1\over 1{-}\del} + 2\del_\rho
        - 2\; {1{-}3\del/2 \over (1{-}\del)^2}\; (\delr{-}\delc) \cr
    & & \quad - {1 \over (1{-}\del)^2}\; {(\delr{-}\delc)^2 \over 4\delg} \cr
     && \quad + 2\delg\; {\del(1{-}3\del/2) \over (1{-}\del)^2}
     \bigg\} \;\; \dgammap \;+ \cr
+ &&\bigg\{{1\over 1{-}\del} - \del_\rho
        + {1{-}2\del \over (1{-}\del)^2}\; (\delr{-}\delc) \cr
    && \quad + {\del \over (1{-}\del)^2}\; {(\delr{-}\delc)^2 \over 4\delg}\cr
     && \quad - \delg\; {\del(1{-}2\del) \over (1{-}\del)^2} 
     \bigg\}\;\; \drho  \;,
\label{eqa:b1}
 \end{eqnarray}
 \begin{eqnarray}
{\delta B_2 \over g/c} &=& \bigg[
        {-} 2\; {1{-}3\del/2 \over (1{-}\del)^2}
        + {1{-}\del \over 2(1{-}\del)^2}\;
          {\delr{-}\delc \over 2\delg} \bigg]\;
        \dgammap \;+ \cr
&+& \bigg[ {1{-}2\del \over (1{-}\del)^2}
        + {\del \over (1{-}\del)^2}\; {\delr{-}\delc \over 2\delg}
        \bigg]\;\; \drho \;,
\label{eqa:b2}
 \end{eqnarray}
  and
 \begin{equation}
\delta B_3 = {1\over 2}\; {1\over 1{-}\del}\;\; \dgammap
        + {1\over 2}\; {\del \over (1{-}\del)^2}\;\;
        \drho \;.
\label{eqa:b3}
 \end{equation}
  where $r$, $\rho$, $c$ and $g$ are distance from the centre, density,
adiabatic sound speed and gravitational acceleration, respectively.
  We have also introduced the following quantities
 \begin{equation}
\del = {l(l{+}1)c^2 \over r^2\omega^2} \;,
\label{eqa:del}
 \end{equation}
  where $l$ is the mode degree, and
 \begin{eqnarray}
\delr &=& {g \over \omega^2c}\; \devdf{}{\tau}
         \log\left({g \over \rho c}\right) \;, \\
\delc &=& {g \over \omega^2c}\; \devdf{}{\tau}
         \log\left({g \over r^2}\right) \;, \\
\delg &=& \left({g \over \omega c}\right)^2 \;.
 \end{eqnarray}
  These are all first order quantities, compared to unity, because well
inside the resonance cavity of the modes the local wavelength is
significantly smaller than the scale of variations of the equilibrium
quantities.

  In order to use the expression for $\delta I$ from Eq.~\ref{eqa:deltai},
it is necessary to replace the term in $(\diff\delta B_0/\diff\tau)$ by
integrating by parts to obtain for $\delta I$;
 \begin{equation}
\delta I = \!\int_{\tau_a}^{\tau_b}\!\! \left[ \delta B_1 E_r^2
        + \left(\delta B_2 {+} \delta B_0\right) \devdf{E_r^2}{\tau}
        + \delta B_3 \devds{E_r^2}{\tau}
        \right]\! \diff\tau .
\end{equation}
   The integration is done only for the region of the ionization zone,
starting at $\tau_a$ and ending at $\tau_b$.
  Because we are restricting our analysis to localized variations, it is
also assumed that the model differences are zero everywhere else.
  This does not affect our result since we will only take those changes in
the frequencies that are not affected by model differences spreading over
regions of size of the order of (or larger than) the local wavelength of
the modes.

  We recall, from asymptotic analysis, that
 \begin{eqnarray}
&&E_r^2 \sim {E_o^2 \over 2}\; \cos(\Lambda) \;,\nonumber \\
&&\devdf{E_r^2}{\tau} \sim - {E_0^2 \over 2}\; 2\omega(1{-}\del)^{1{/}2}
        \; \sin(\Lambda) \;, \\
&&\devds{E_r^2}{\tau} \sim - {E_0^2 \over 2}\; 4\omega^2(1{-}\del)
        \; \cos(\Lambda) \;.\nonumber
\end{eqnarray}
  The argument of the trigonometric functions is 
 \begin{equation}
\Lambda(\tau) \equiv 2 \left[
   \omega \int_0^\tau \left(1{-}\del\right)^{1/2}\diff\tau + \phi \right] \;.
 \end{equation}
  After replacing these expressions in the equation for $\delta I$, we
find
 \begin{eqnarray}
{2 \over \omega^2E_o^2}\; \delta I &\sim&  \int_{\tau_a}^{\tau_b}
        \Bigg\{ \left[ {\delta B_1 \over \omega^2}
                  - 4(1{-}\del)\delta B_3\right] \cos\Lambda \cr
        &&{-} 2(1{-}\del)^{1/2} {\delta B_2 {+} \delta B_0 \over \omega}
                  \sin\Lambda \Bigg\} \diff\tau \;.
\label{eqa:delta-i}
 \end{eqnarray}
  This expression gives the variational principle for perturbations in the
frequencies due to small changes in the structure, as described by $\delta
B_i$.

  The next step is to establish what is the effect on the structure of
the ionization zone for helium, relative to a model where such a
localized effect is not present.
  In particular, we need to estimate how $\Gamma_1$, $P$ and $\rho$ are
changed from being slowly varying functions of depth to the actual values they
have when the second ionization of helium occurs.
  The difference will produce the $\delta(\Gamma_1 P)$ and $\delta\rho$
responsible for changing the frequencies, as given in
Eqs~(\ref{eqa:b0}-\ref{eqa:b3}).
  That will allow us to calculate an expression for the characteristic
signal we want to isolate in the frequencies.

  In order to find an expression for the signal we shall first consider
that the changes are dominated by $\Gamma_1$.
  In doing so, we adopt here a different approach from \cite{monteiro96t},
who consider that the dominant contribution could be isolated in the
derivative of the sound speed.
  We do so because the effect of the ionization is better represented as a
`bump' in $\Gamma_1$ (see Figs~\ref{fig:gamma1} and \ref{fig:delta-gamma1}),
extending over a localized region of the Sun.
  Therefore we retain the terms for $\delta\Gamma_1$, and neglect, as a
first approximation, the contributions from $\delta\rho$ and $\delta P$.
  In doing so we assume that the changes in the sound speed are mainly due
to the changes in the adiabatic exponent.

  Now, relating $\delta I$ to the change in the eigenvalue $\delta\omega$
(and using Eq.~\ref{eqa:var_princ}) it follows that
 \begin{eqnarray}
\left[\delta\omega\right]_{\Gamma_1} &\equiv&
     {\left[\delta I\right]_{\Gamma_1} \over \omega\taut E_o^2} \cr
&\sim& {\omega \over 2 \taut}
   \int_{\tau_a}^{\tau_b} \left(f_c \cos\Lambda + f_s \sin\Lambda\right)
     {\delta\Gamma_1 \over \Gamma_1}\; \diff\tau \;,
\label{eqa:sig-gamma1}
 \end{eqnarray}
  where $f_s$ and $f_c$ are functions obtained from adding the
coefficients of $\delta\Gamma_1$ in the expressions of $\delta B_0$,
$\delta B_1$, $\delta B_2$ and $\delta B_3$ (see Eq.~\ref{eqa:delta-i} and
Eqs~\ref{eqa:b0}-\ref{eqa:b3}).

  At this point we introduce an approximate description of the effect of
the second ionization of helium on the adiabatic exponent.
  As represented in Fig.~\ref{fig:delta-gamma1}b, we adopt a prescription
where the `bump' is approximately described by its half width $\beta$ and
height $\deltad{\equiv}(\delta\Gamma_1/\Gamma_1)_{\taud}$, with the
maximum located at $\taud$.
  This corresponds to considering the following approximating simple
expression for $\delta\Gamma_1$:
 \begin{equation}
{\delta\Gamma_1 \over \Gamma_1} \equiv \deltad
\cases{
   \dis \left( 1 {+} {\tau{-}\tau_d \over\beta} \right)
   & ; $\tau_d {-}(1{-}\alpha)\beta \le \tau \le \tau_d$ \cr
      \cr
   \dis \left( 1 {-} {\tau{-}\tau_d \over\beta} \right)
   & ; $\tau_d \le \tau \le \tau_d{+}(1{+}\alpha)\beta$ \cr
\cr
\quad 0 & ; elsewhere. \cr}
 \end{equation}
   The region of the ionization zone starts at
$\tau_a{=}\taud{-}(1{-}\alpha)\beta$ and finishes for
$\tau_b{=}\taud{+}(1{+}\alpha)\beta$, giving that
$\tau_b{-}\tau_a{=}2\beta$ is the width.
  The parameter $\alpha$ represents the asymmetry of the bump, and for a
first order analysis it does not affect the result.

  We further consider that the functions $f_s$ and $f_c$ are slowly
varying functions of the structure when compared with the size of the
ionization zone $({\sim} 2\beta)$, and so their derivatives can be ignored
in the integration.
  Using this approximation we may integrate Eq.~(\ref{eqa:sig-gamma1})
finding that
 \begin{eqnarray}
\left[\delta\omega\right]_{\Gamma_1} &\sim& 
      {\omega \over 2\taut}\; \beta\, \deltad
      \left\{{\sin\left[\omega\beta(1{-}\del)^{1/2}\right]\over
   \omega\beta(1{-}\del)^{1/2}}\right\}^2 \cr
 && \qquad\times \Big(f_c\; \cos\Lambdad + f_s\; \sin\Lambdad\Big) \;. 
\label{eqb:sig_full}
 \end{eqnarray}
  All quantities are now evaluated at $\tau{=}\taud$.

  Taking the dominant contributions (in terms of powers of $\omega$ and
derivatives of the reference structure -- see CDMT for details) of the
functions $f_c$ and $f_s$ (Eq.~\ref{eqa:delta-i}), we can finally write
the signal as being
 \begin{equation}
[\delta\omega]_{\Gamma_1} \sim {3 \deltad \over 2 \taut} \;
   {1{-}2\del/3 \over 1{-}\del}
   \; {\sin^2\left[\omega\beta(1{-}\del)^{1/2}\right] \over
            \omega\beta(1{-}\del)}
   \; \cos\Lambdad \;.
\label{eqb:signal}
\end{equation}

  This is the expression that describes the `additional' contribution
to the frequencies of oscillation $\omega_{nl}$ if the region of
the second ionization of helium is present.
   By assuming that we have
\begin{equation}
\omega_{nl} \equiv [\omega_{nl}]_{\rm smooth} + 
[\delta\omega_{nl}]_{\Gamma_1}\;,
\end{equation}
it is now possible to try removing the smooth component 
$[\omega_{nl}]_{\rm smooth}$,
by adjusting the frequencies to the expression we have found for
the `periodic' component $[\delta\omega_{nl}]_{\Gamma_1}$.
In doing so the parameters describing the structure of the Sun at the
location $\taud$ are determined.

\label{lastpage}
\end{document}